\documentclass[final,5p,times,twocolumn]{elsarticle} 

\usepackage{amssymb}
\usepackage{amsmath}

\usepackage{subcaption}
\graphicspath{{./}}

\journal{Nuclear Instruments & Methods in Physics Research, Section A}

\begin{document}
\begin{frontmatter}


\title{A Low Power Monolithic Active Pixel Sensor Prototype for the STCF Inner Tracker} 


\author[1,2]{Dongwei Xuan} 
\author[1,2]{Ruiyang Zhang}
\author[1,2]{Jiajun Qin}
\author[1,2]{Hao Han}
\author[1,2]{Xinyu Bin}
\author[1,2]{Zihan Xu}
\author[1,2]{Lei Zhao}
\author[1,2]{Jianbei Liu}
\author[3]{Liang Zhang}
\author[3]{Anqing Wang}
\author[3]{Aodong Song}
\author[4]{Xiangming Sun}
\author[4]{Le Xiao}
\author[1,2]{Lailin Xu \corref{cor1}}
\affiliation[1]{organization={State Key Laboratory of Particle Detection and Electronics},
            addressline={University of Science and Technology of China}, 
            city={Hefei},
            postcode={230026}, 
            country={China}}
\affiliation[2]{organization={Department of Modern Physics},
            addressline={University of Science and Technology of China}, 
            city={Hefei},
            postcode={230026}, 
            country={China}}
\affiliation[3]{organization={Institute of Frontier and Interdisciplinary Science and Key Laboratory of Particle Physics and Particle Irradiation (MOE)},
            addressline={Shandong University}, 
            city={Qingdao},
            postcode={266237}, 
            country={China}}
\affiliation[4]{organization={PLAC, Key Laboratory of Quark and Lepton Physics (MOE)},
            addressline={Central China Normal University}, 
            city={Wuhan},
            postcode={430079}, 
            country={China}}
            
\cortext[cor1]{Corresponding author: Lailin Xu (lailinxu@ustc.edu.cn)}

\begin{abstract}
The Super Tau-Charm Facility (STCF) is a proposed $e^+e^-$ collider with a peak luminosity 100 times higher than that of the present tau-charm factory. The inner tracker (ITK) of STCF should feature a low material budget and high readout speed. Under these requirements, the monolithic active pixel sensor (MAPS) is considered as a promising candidate for the ITK. To minimize the power consumption of MAPS (for low material budget), larger-size sensors are proposed to reduce the scale of the readout circuitry while preserving the required position resolution. Multiple sensors with varying dimensions and structures were designed and integrated in several prototype chips for performance comparison, fabricated in a 180~nm CIS process. The in-pixel readout circuit can also provide time of arrival (ToA) and time-over-threshold (ToT) of the hit signal, with a least significant bit (LSB) of 50~ns. The peripheral readout circuit performs operations including timestamp correction, data aggregation, caching, framing, 8b/10b encoding, and serialization. According to simulation, the power consumption for a full-scale chip is about 55.7~mW/cm$^2$. Preliminary measurements have been conducted on the prototype chips.
\end{abstract}

\begin{graphicalabstract}
\end{graphicalabstract}

\begin{highlights}
\item A MAPS chip with dedicated low-power digital readout architecture was designed for the STCF inner tracker detector with required spatial and timing resolutions, focusing on minimizing digital power consumption across the pixel array.
\item Laboratory tests were conducted to validate the prototype chip's functionality and evaluate its performance.
\end{highlights}

\begin{keyword}
STCF, MAPS, pixel readout circuit, low power consumption

\end{keyword}

\end{frontmatter}



\section{Introduction}
\label{sec1}
The Super Tau-Charm Facility (STCF) is a proposed next-generation high-luminosity $e^+e^-$ collider designed to operate in the tau-charm energy region of 2 to 7 GeV \cite{achasov2024stcf}. The peak luminosity is over $0.5 \times 10^{35}\ \mathrm{cm}^{-2}\cdot \mathrm{s}^{-1}$ at the optimized center-of-mass energy of 4 GeV. To meet the STCF's physics targets and maximize its physics potential, the detector must achieve the stringent performance requirements at high luminosity. As the sub-detector closest to the beam pipe, the inner tracker (ITK) has the primary task of providing sufficient detection efficiency for low momentum particles: the detection efficiency should exceed 90\% for particles with a momentum of about 0.1~GeV/c. The ITK thus needs an ultra-low material budget (<0.3\% $X_0$ per layer) to reduce the multi-scattering effect for low-momentum charged particles.

The monolithic active pixel sensor (MAPS) has been considered as a promising candidate for the ITK due to its low material budget and high readout speed. By integrating both the sensor and readout circuit on the same silicon wafer, MAPS achieves a significantly reduced material budget compared to traditional hybrid pixel detectors. The MAPS-based ITK consists of three cylindrical layers. The innermost layer (at a radius of 36~mm) is required to withstand the highest hit rate (1~MHz/cm$^2$) and radiation environment, with Total Ionizing Dose (TID) of 10~kGy/year and Non-Ionizing Energy Loss (NIEL) of 1$\times$$10^{11}$ $n$/cm$^2$/year \cite{fang2024study}. To further mitigate the pileup effects, the timing resolution of STCF MAPS needs to reach at least 50~ns initially, with the goal of ultimately achieving 20~ns. While precise vertex reconstruction is not currently required for the ITK, its spatial resolution is intentionally relaxed to 100~$\mu$m - a target well within the demonstrated capability of MAPS. 

The dominant challenge for the STCF MAPS design is to satisfy the extremly low material budget requirement. This constraint necessitates low-power readout circuitry (<~100~mW/cm$^2$), thereby reducing the mass of cooling systems and cables.
It is even more challenging for the MAPS with timing resolution requirements. Because the established method for measuring ToA requires distributing timestamps to each pixel, generating substantial digital power consumption. For instance, the timestamp distribution in TJ-Monopix2 consumes 80~mW/cm$^2$ with a 25~ns bin size, accounting for about $50\%$ of the total chip power consumption~\cite{moustakas2021design}. 

Consequently, the STCF MAPS prototype chip design primarily focuses on minimizing the timestamp distribution power consumption. This power consumption is proportional to $F_{\mathrm{clk}} \cdot N_{\mathrm{col}} \cdot C_{\mathrm{par}} \cdot V_{\mathrm{DD}}^2$, where $F_{\mathrm{clk}}$ is the timestamp clock frequency, $N_{\mathrm{col}}$ is the number of pixel array columns per unit width, $C_{\mathrm{par}}$ is the parasitic capacitance per column and $V_{\mathrm{DD}}^2$ is the digital supply voltage. Thus multiple large-size sensors with varying dimensions and structures were implemented. By extending the pixel pitch in the horizontal direction up to five times that of TJ-Monopix2, the $N_{\mathrm{col}}$ can be significantly reduced, leading to a substantial decrease in power consumption needed for the timestamp distribution. Moreover, for column-based readout architectures, reducing $N_{\mathrm{col}}$ directly scales down the peripheral readout circuit, consequently reducing its power consumption.

\section{Sensor layout}
\label{sec2} 

\begin{figure}[]
\centering
\subfloat[(a)]{
    \includegraphics[width=0.08\linewidth]{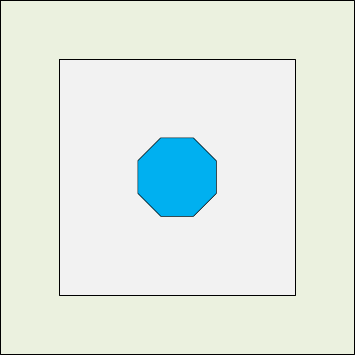} 
}
\hfill
\subfloat[(b)]{
    \includegraphics[width=0.48\linewidth]{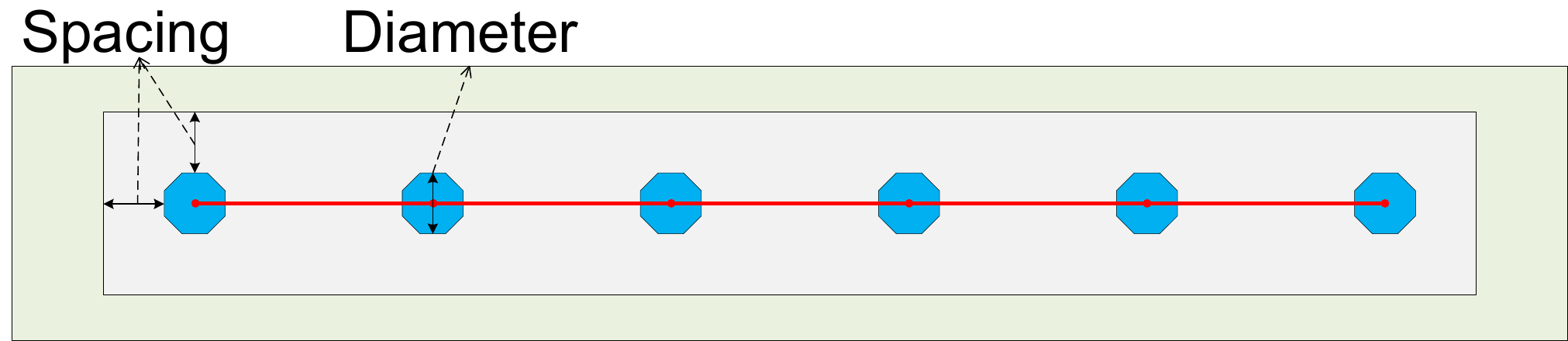}
}
\hfill
\subfloat[(c)]{
    \includegraphics[width=0.24\linewidth]{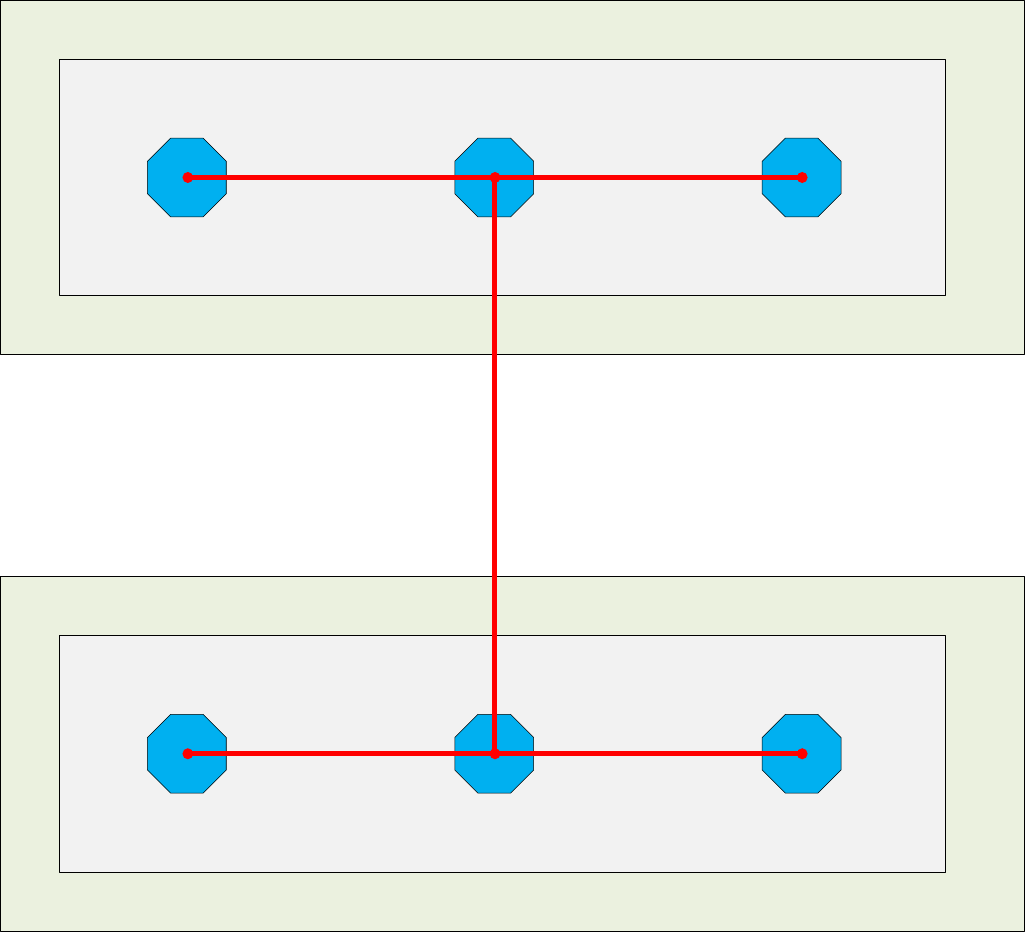}
}

\subfloat{
    \includegraphics[width=0.16\linewidth]{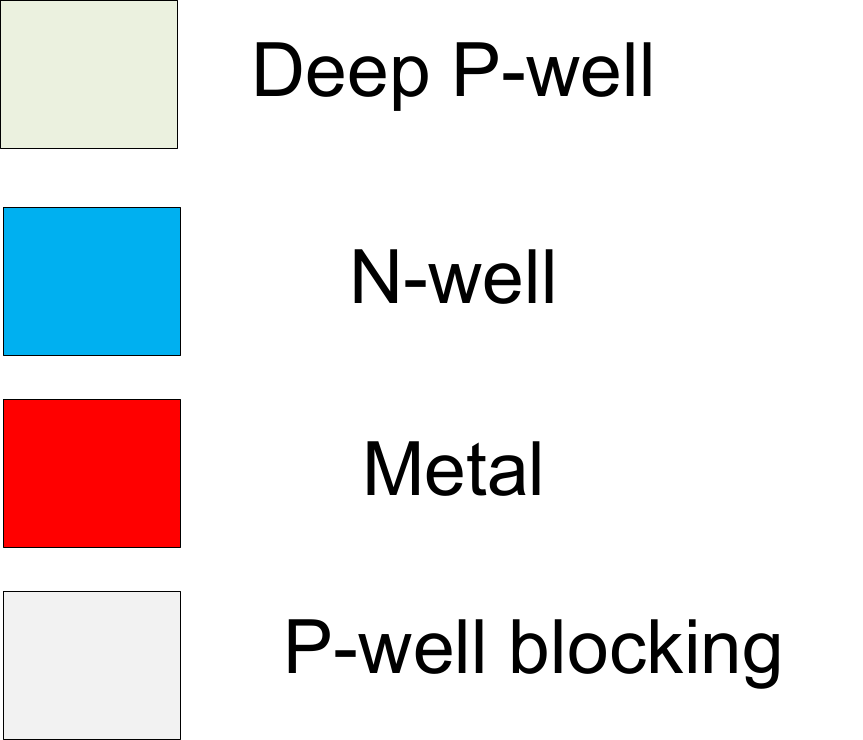} 
}
\hfill
\subfloat[(d)]{
    \includegraphics[width=0.48\linewidth]{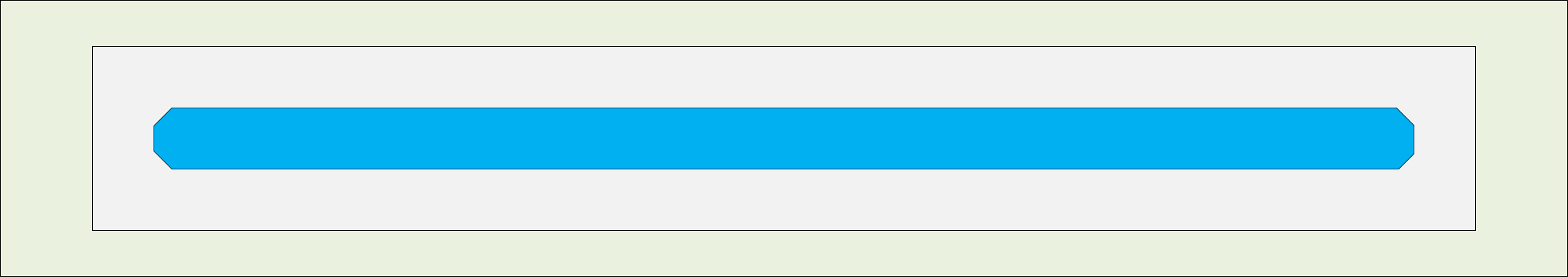}
}
\hfill
\subfloat[(e)]{
    \includegraphics[width=0.24\linewidth]{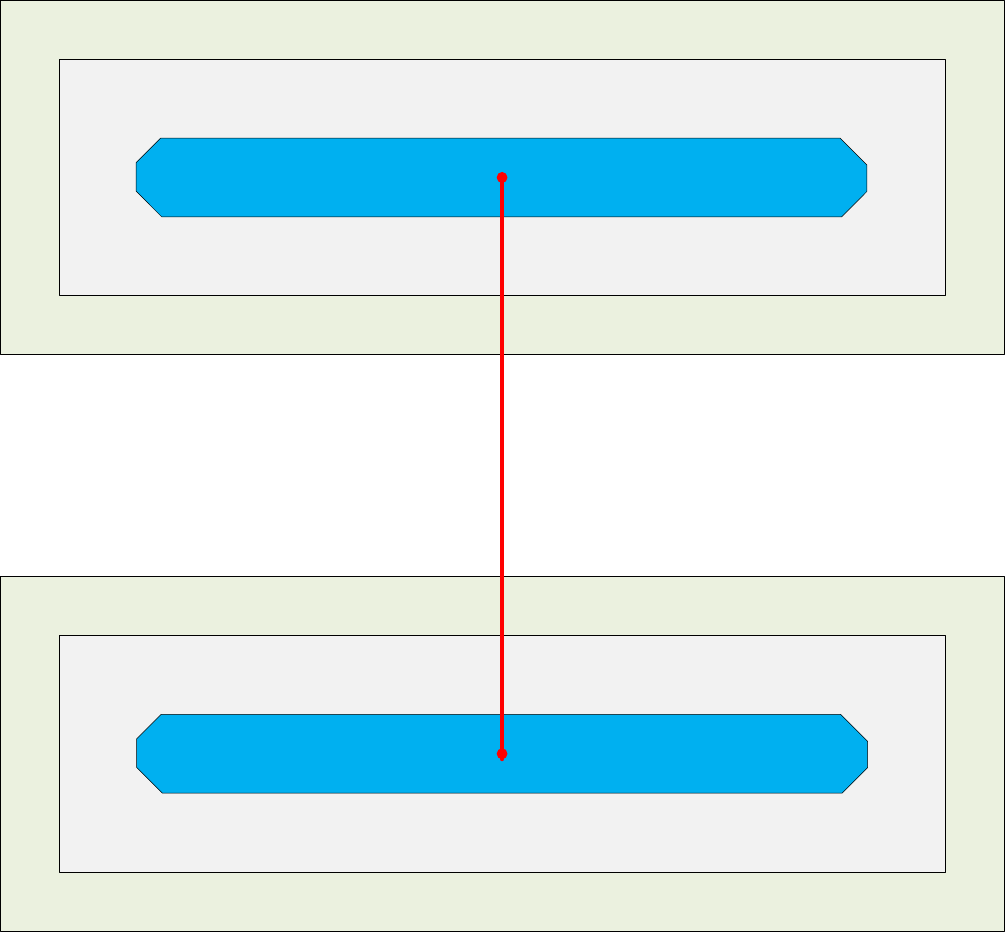}
}
\hfill
\caption{Sensor structures: (a) small sensor as a reference, (b) metal-connect large pixel with single row, (c) metal-connect large pixel with double rows, (d) active-connect large pixel with single row, (e) active-connect large pixel with double rows}\label{fig2}
\end {figure}

A variety of sensors with different structures have been designed to verify and compare their performance. The diagrams of these sensors are shown in Figure~\ref{fig2}. Among these sensors, the ``active-connect'' type is obtained by stretching the octagonal-shaped n-well collection electrode. The height of the collector electrode is 2~$\mu$m, but its widths are stretched to 140~$\mu$m (Sensor D) and 70~$\mu$m (Sensor E), respectively. The distance from each side of the collector to the surrounding deep P-well is 2~$\mu$m. While the ``metal-connect'' type of sensor is obtained by connecting multiple octagonal-shaped collection electrodes using a metal layer with a diameter of 2~$\mu$m (Sensor B and Sensor C). 

All types of sensors have been modeled and simulated in TCAD. Compared to metal-connect sensors, active-connect sensors exhibit relatively larger depletion regions, which leads to a faster charge collection speed and also reduced charge sharing effects. Taking Sensor D as an example, in an extreme condition where a particle strikes the center of four neighbouring pixels, the charge collection time for one sensor is about 35~ns. This value increases to 55~ns for metal-connect sensors with the same dimensions (Sensor B). However, active-connect sensors also display relatively larger parasitic capacitance, which negatively impacts the front-end's charge gain and noise performance. Therefore, applying a negative bias to the substrate becomes crucial for these large-sized sensors, in contrast to conventional small-sized pixels. According to TCAD simulations, when the substrate bias voltage is reduced from 0~V to -6~V, the detector capacitance of Sensor D decreases from 69~fF to 42~fF. In comparison, this variation ranges from 28~fF to 23~fF for Sensor B.

\section{Prototype architecture}
\label{sec3}  
\begin{figure}
    \centering
    \includegraphics[width=0.75\linewidth]{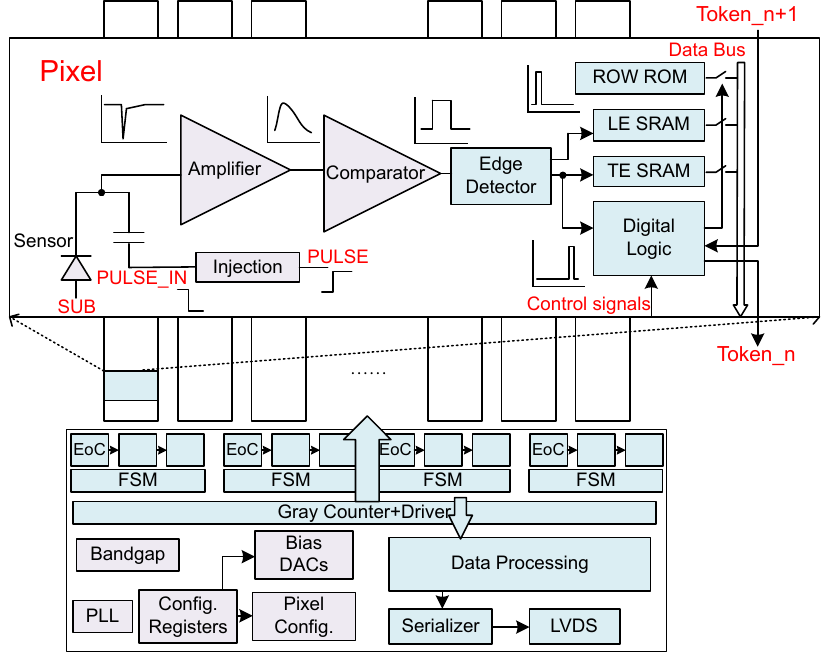}
    \caption{Block diagram of the MAPS}
    \label{fig1}
\end{figure}

Multiple prototype chips with these types of sensors have been designed and fabricated in a 180 nm CMOS image sensor process, featuring a high-resistivity(>1~$k\Omega\cdot cm$) epitaxial layer with a thickness of about 20~$\mu$m. One of the prototype chips, named CharTPix-TJ-v0.1, incorporates two sectors: a 7~$\times$~60 array with Sensor B and an 8~$\times$~60 array with Sensor D, both featuring identical pixel dimensions of 31~$\mu$m~$\times$~170~$\mu$m. The block diagram of these prototypes is shown in Figure~\ref{fig1}. CharTPix-TJ-v0.1 mainly consists of a pixel array and peripheral circuits. Each pixel cell integrates a front-end circuit, priority readout logic, SRAM and ROM. The peripheral circuits include a DAC module, a bandgap, a readout circuit, a phase-locked loop (PLL), a serializer and a LVDS module. Prototype chips incorporating other sensor types also adopt a similar architecture.

\subsection{Front-end circuit}
\label{subsec2}

\begin{figure}
    \centering
    \subfloat[(a)]{
    \includegraphics[width=0.5\linewidth]{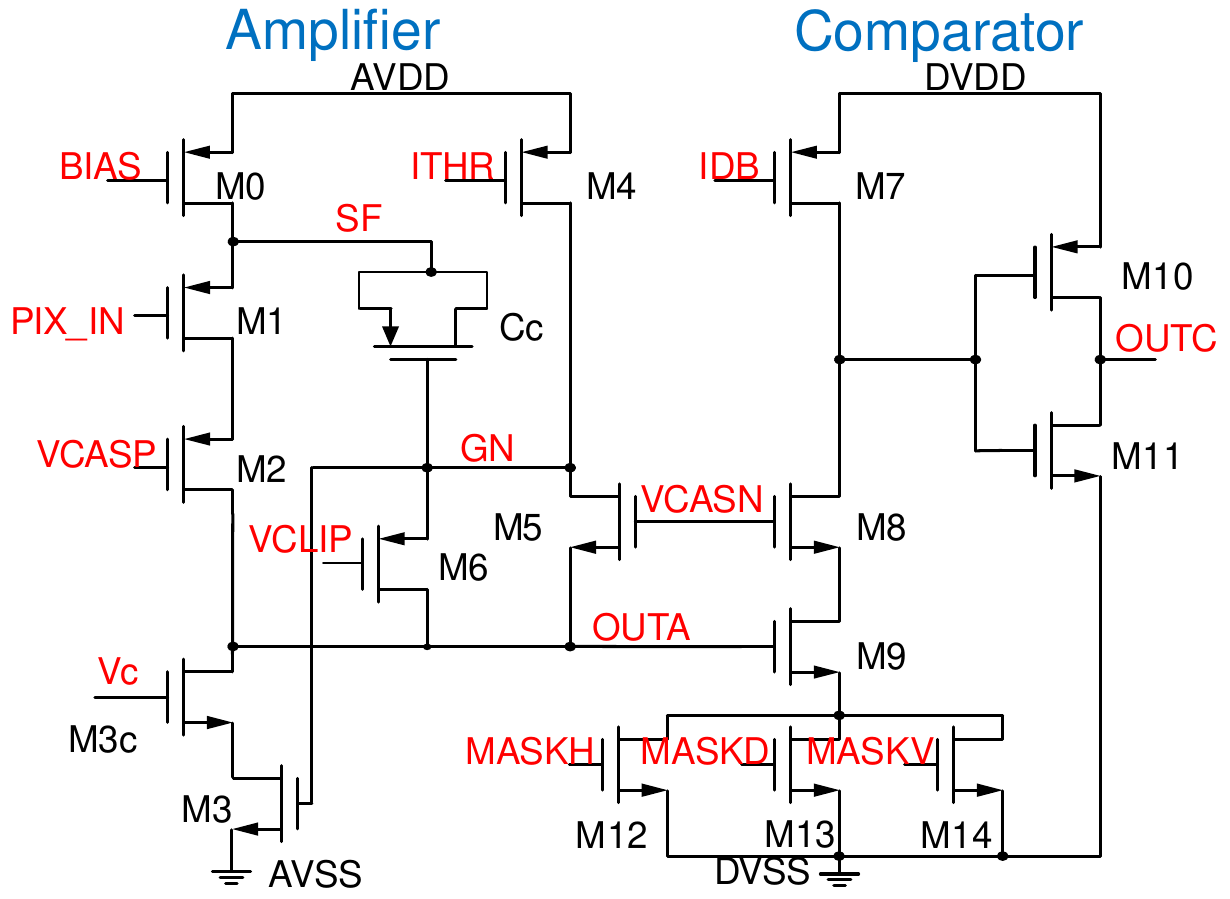}\label{Fig.3(a)}
}

    \subfloat[(b)]{
    \includegraphics[width=0.6\linewidth]{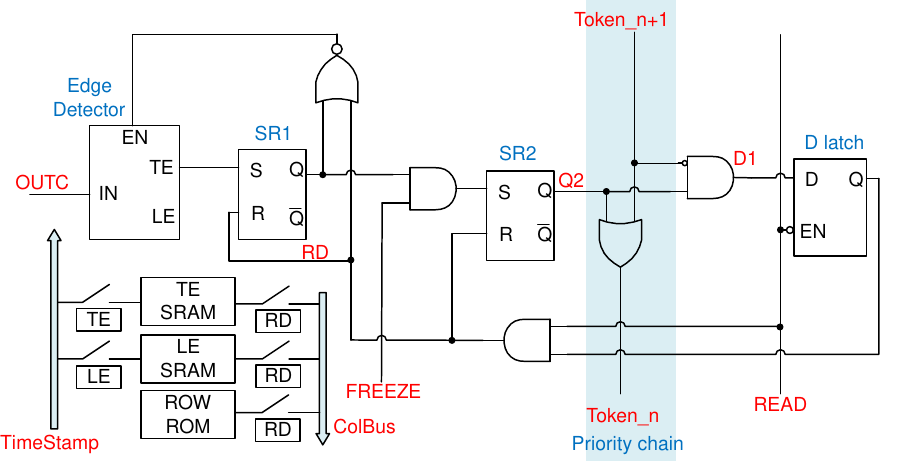} \label{Fig.3(b)}
}
\caption{(a) Pixel front-end circuit, (b) ``column-drain'' readout logic }
\end{figure}

The front-end circuit employs an open-loop architecture derived from the ALPIDE~\cite{kim2016front} chip developed for the ALICE experiment as illustrated in Figure \ref{Fig.3(a)}. The increased sensor size leads to higher input capacitance, thereby imposing stricter requirements on the front-end circuit's voltage gain. Several optimizations were implemented to increase the voltage gain, including: enlarging $C_c$ (>1~pF), raising the bias current BIAS, increasing the gate length of M3, and adding a cascode transistor M3c similar to the MALTA2~\cite{malta2} design. Meanwhile, the higher BIAS reduces the peaking time of the analog output and improves the discriminator's response speed. By taking the simulation result of sensor B as a representative case, it exhibits a time walk of less than 320~ns and a charge threshold of 314~$e^-$, along with an equivalent noise charge (ENC) of 11~$e^-$.

\subsection{Digital readout circuit}
\label{subsec3}
The priority readout logic in the pixel, designated as ``column-drain'' architecture~\cite{peric2006fei3}, selectively transfers data exclusively from triggered pixels to peripheral readout circuits at 10~MHz per column per hit event. The readout circuit diagram is shown in Figure~\ref{Fig.3(b)}. The edge detector captures both the leading edge (LE) and the trailing edge (TE) of the discriminator output, storing the corresponding 8-bit timestamps in SRAM. Additionally, the trailing edge asynchronously sets both SR latches and asserts the ``Token'' signal. The ``Token'' propagates to peripheral circuits through a column-wise OR gate chain. When the state machine detects this asserted signal, it initiates readout operations by generating a READ pulse. Among all triggered pixels, the highest-row (highest-priority) pixel's data, including LE, TE and row address, is the first to be read out to the End-of-Column and subsequently output to the peripheral readout circuit.

The peripheral readout circuit performs operations including timestamp correction, data aggregation, caching, framing, 8b/10b encoding, and serialization. Two 800~Mbps LVDS channels are used for data transmission, supporting an event rate up to 8.7~MHz/cm$^2$, which is more than ten times of the average event rate of the innermost layer of the STCF ITK.

To meet the stringent power consumption target, detailed power simulation and analysis were conducted under the hypothesis of scaling the prototype design to a dimension of 2~cm~$\times$~2~cm. The average power consumption of the chip is 55.7~mW/cm$^2$ (Sensor D), where the matrix analog power is approximately 26~mW/cm$^2$. The timestamp distribution consumes 12.2~mW/cm$^2$, representing a significant reduction compared to other chips employing smaller pixel dimensions. A dedicated peripheral readout circuit matching this full-scale array dimensions was designed. Following the completion of place-and-route, power simulations were performed using PrimeTime~PX. The peripheral digital circuit consumes only 23.5~mW at the design hit rate, due to the dedicated optimization described previously.

\section{Experimental results}
\label{sec4}
\begin{figure}
    \centering
    \includegraphics[width=0.6\linewidth]{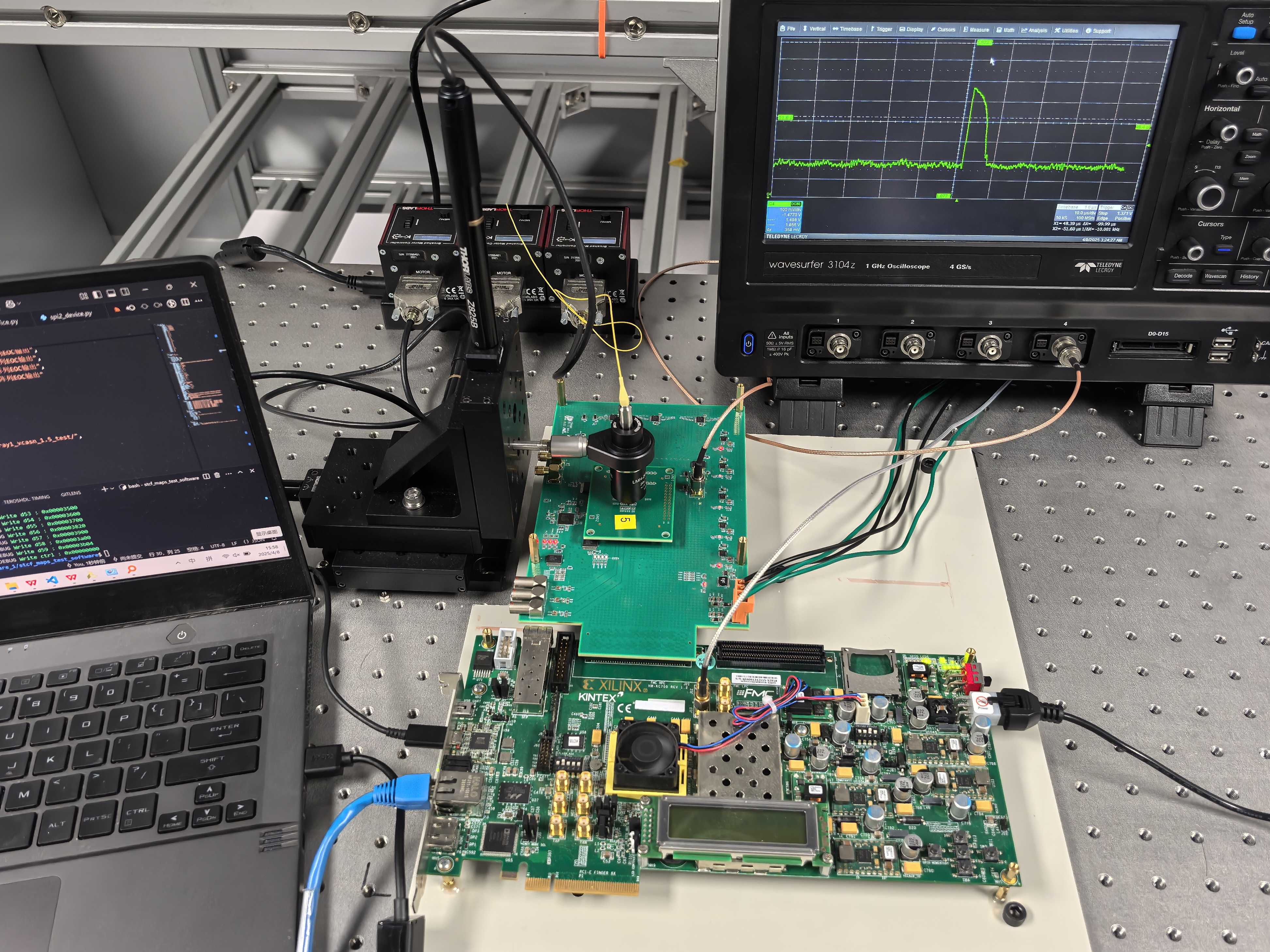}
    \caption{Test system for MAPS prototypes}
    \label{fig:test_system}
\end{figure}

The test system is shown in Figure~\ref{fig:test_system}. It includes an on-chip carrier PCB, a universal mother board, an FPGA board and a PC. The FPGA communicates with the PC via a 1~Gb/s Ethernet link. This MAPS DAQ is based on the IPbus framework~\cite{larrea2015ipbus}. Preliminary measurements were performed on the prototype chips integrating both Sensor B and Sensor D.

\subsection{Electrical test}

\begin{figure}
    \centering
    \subfloat[(a)]{
        \includegraphics[width=0.42\linewidth]{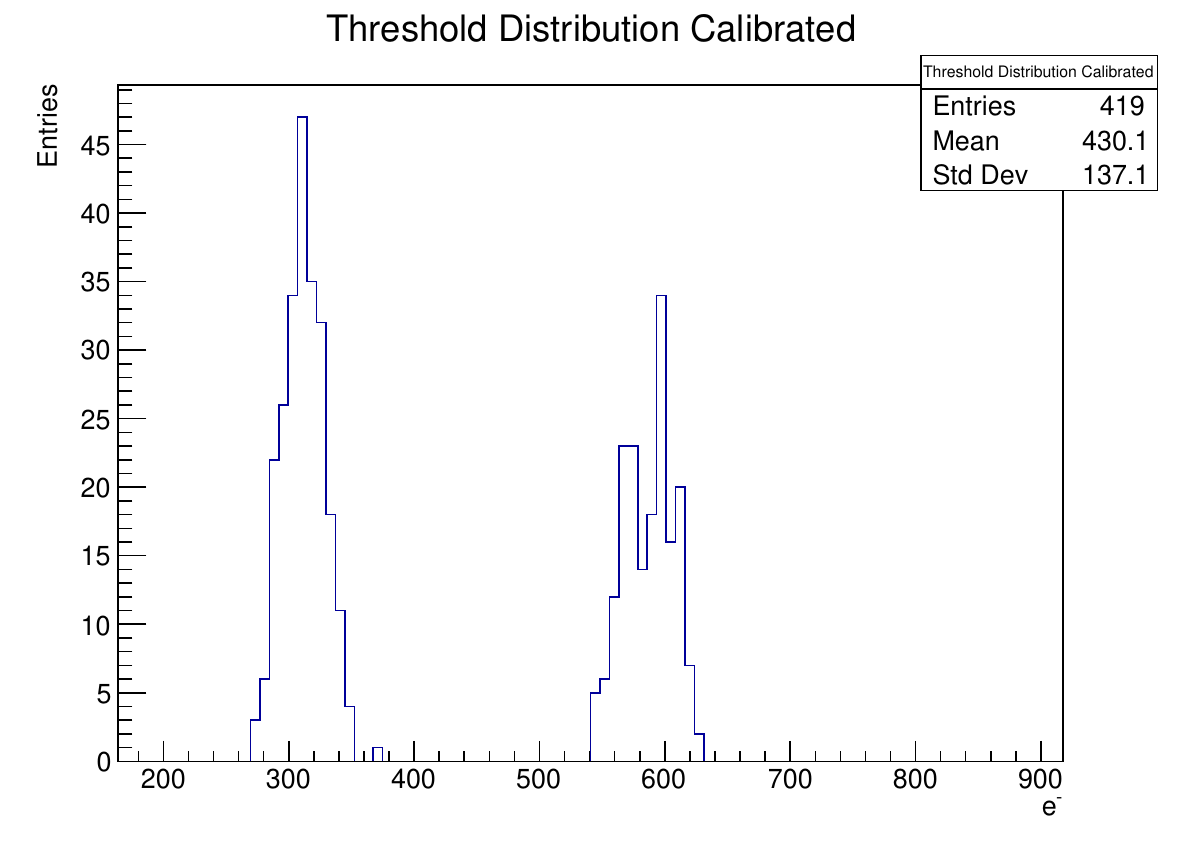}\label{Fig.5(a)}
    }
    \subfloat[(b)]{
        \includegraphics[width=0.42\linewidth]{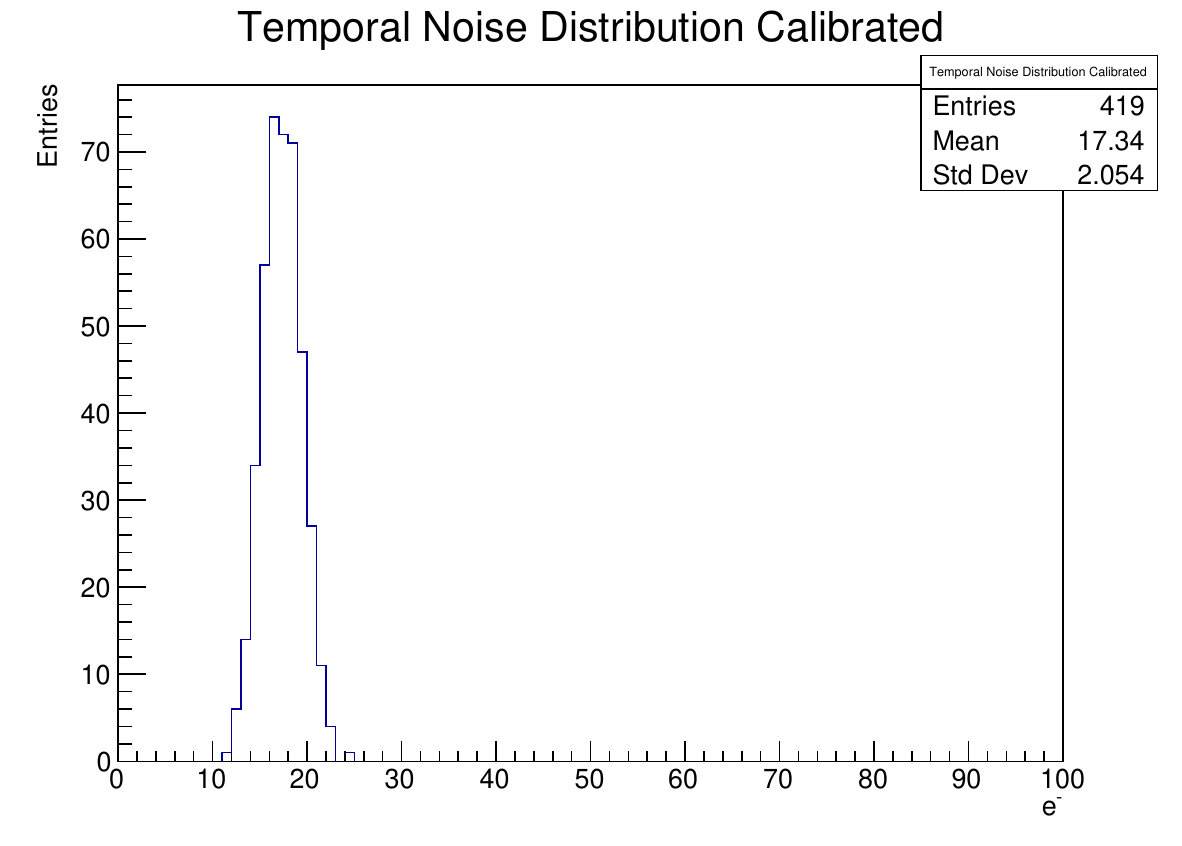} \label{Fig.5(b)}
    }

    \subfloat[(c)]{
        \includegraphics[width=0.42\linewidth]{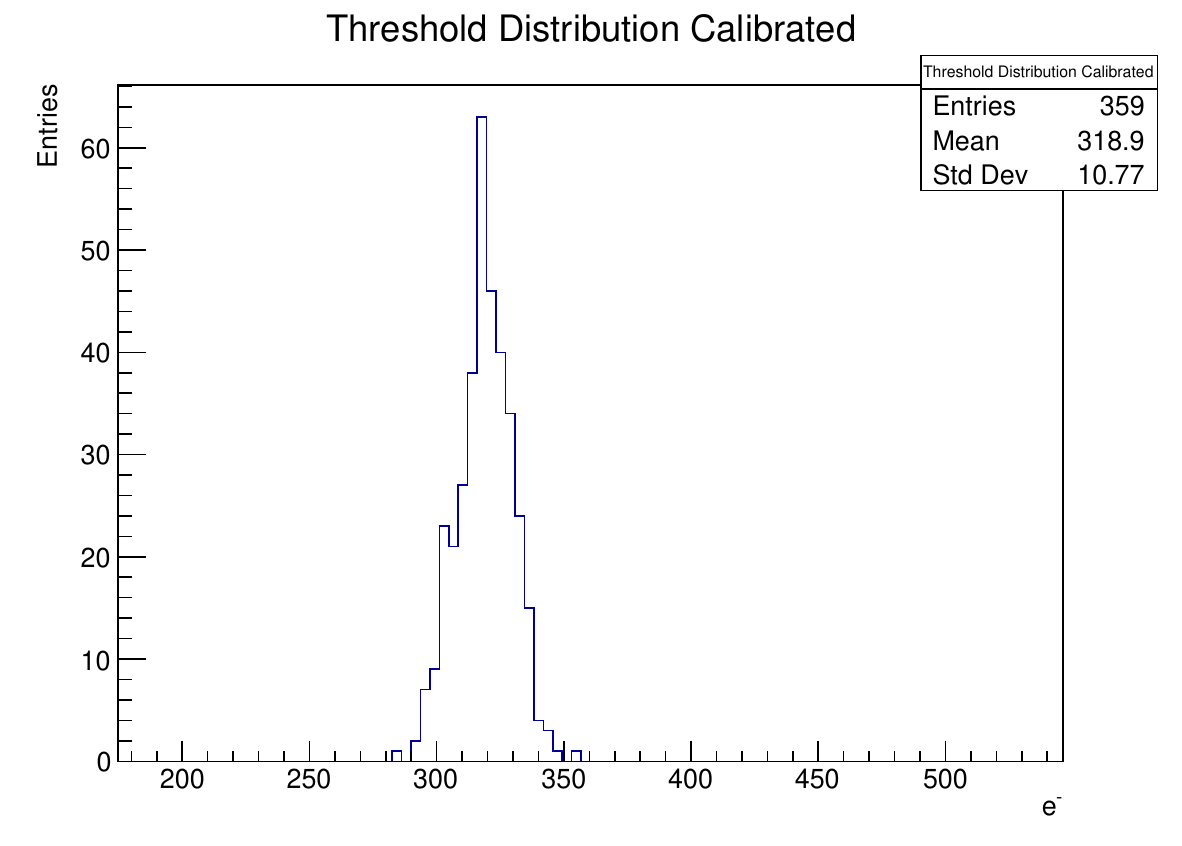}\label{Fig.5(c)}
    }
    \subfloat[(d)]{
        \includegraphics[width=0.42\linewidth]{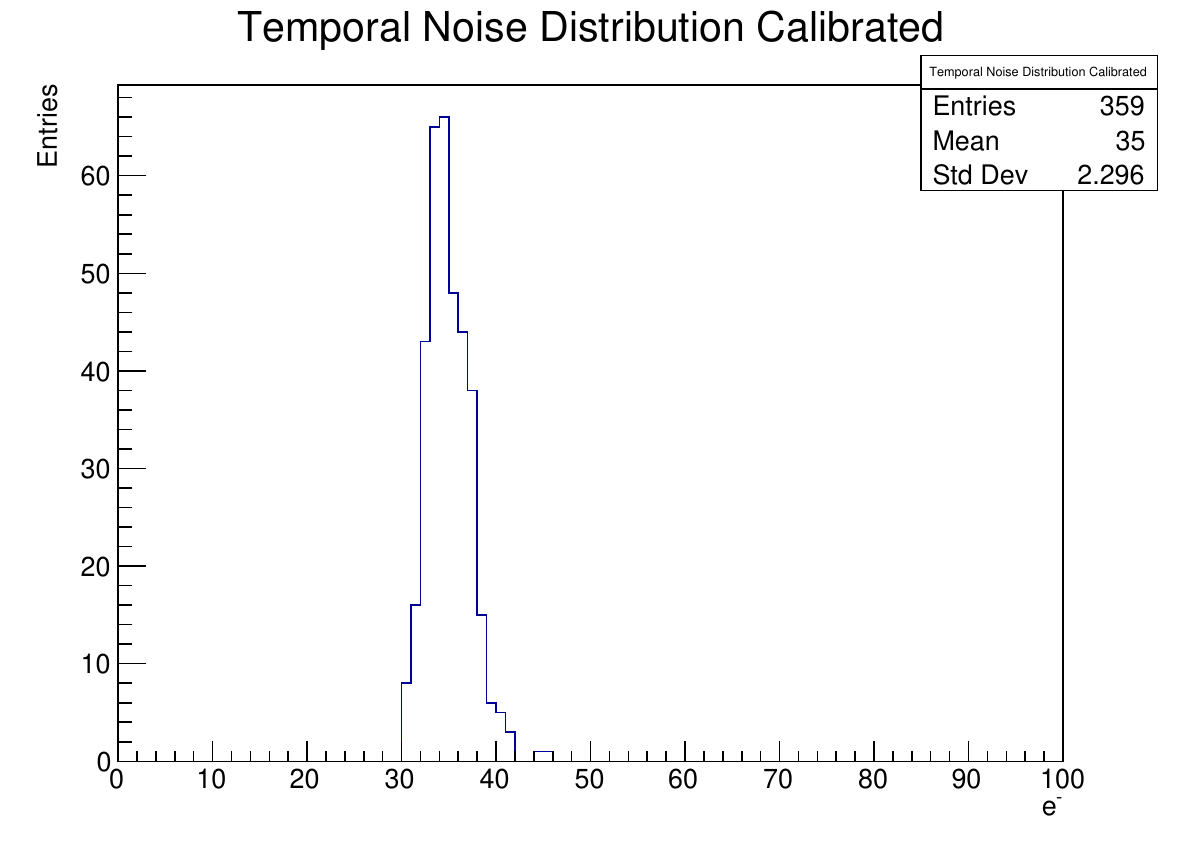} \label{Fig.5(d)}
    }
\caption{(a) Threshold distribution of Sensor B, (b) ENC distribution of Sensor B, (c) Threshold distribution of Sensor D, (d) ENC distribution of Sensor D}
\label{Fig.6}
\end{figure}

\begin{figure}
    \centering
    \subfloat[(a)]{
        \includegraphics[width=0.4\linewidth]{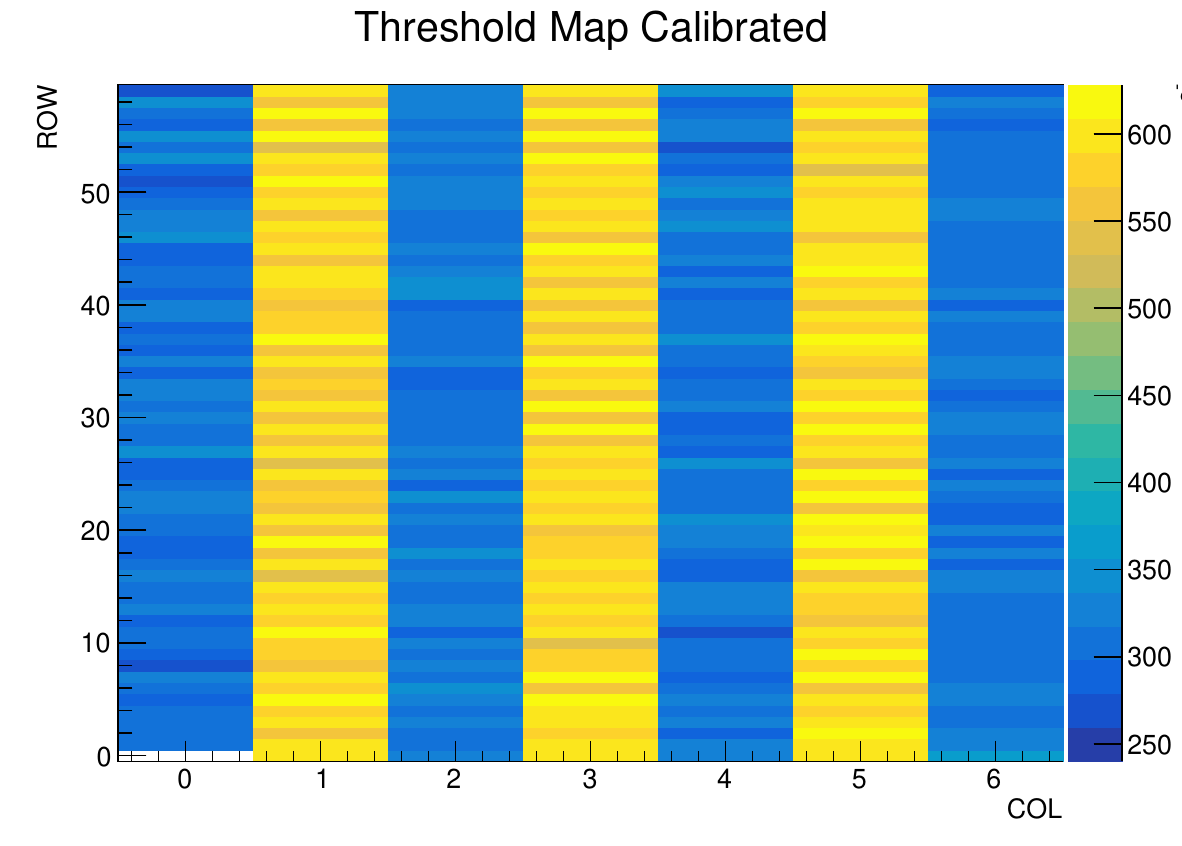} \label{Fig.7(b)}
    }
    \subfloat[(b)]{
        \includegraphics[width=0.4\linewidth]{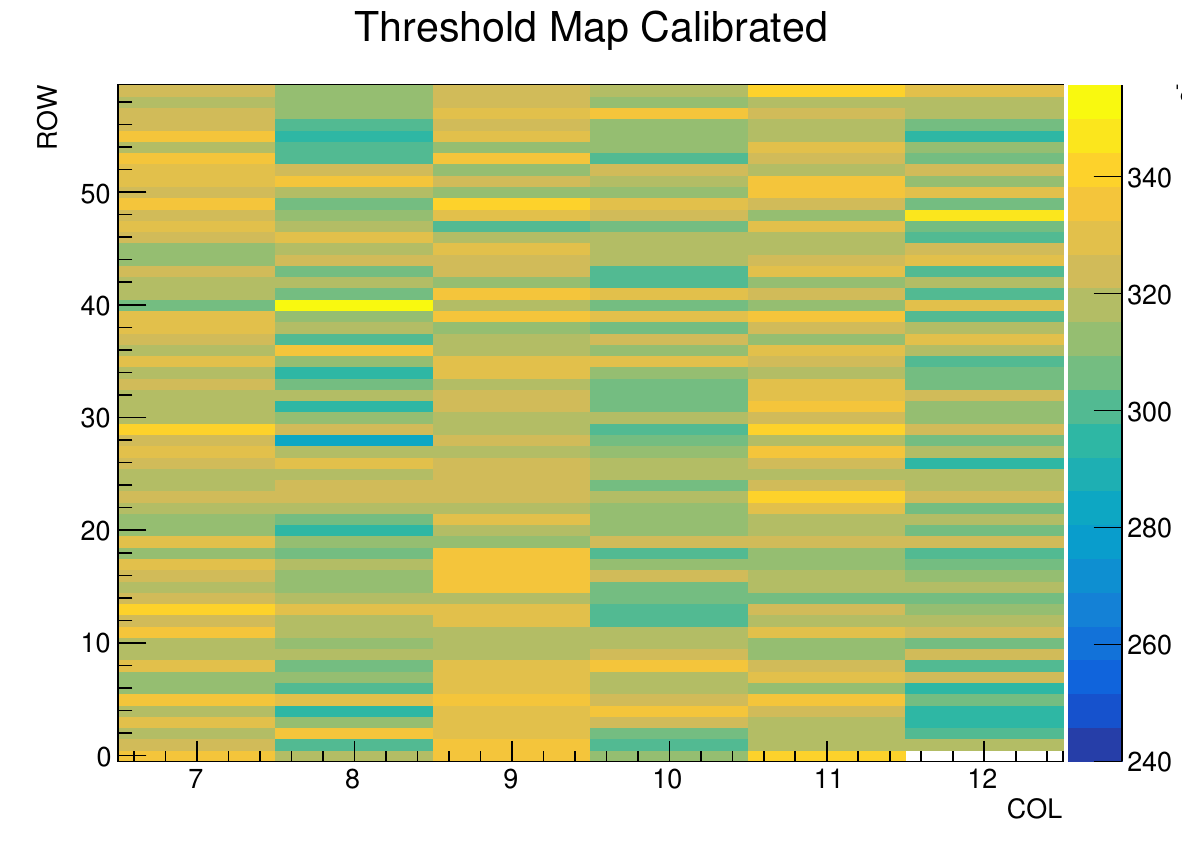}\label{Fig.7(a)}
    }

\caption{(a) Threshold map of Sensor B, (b) Threshold map of Sensor D}
\label{Fig.7}
\end{figure}

Electrical test utilizes the charge injection circuit integrated in each pixel to evaluate the front-end circuit performance and verify the readout functionality. By sweeping the injected charge at the front-end input, the pixel discriminator's flip probability is measured to generate S-curves, which are subsequently fitted with error functions. This fitting process extracts the threshold charge and equivalent noise charge (ENC). Figure~\ref{Fig.6} illustrates the threshold and noise distributions across pixel arrays for both Sensor B and Sensor D under a -6~V substrate bias, after optimizing the settings of the front-end. For Sensor D, the average threshold is 319~$e^-$ with a dispersion of 10.8~$e^-$ . Figure~\ref{Fig.7(a)} displays the two-dimensional threshold map of Sensor D, showing a relatively uniform distribution.

In contrast, Sensor B exhibits column-wise odd-even inconsistency in its threshold distribution (Figure~\ref{Fig.5(a)}). According to the extracted parasitic parameters and the simulation, this effect is hypothesized to originate from the parasitic capacitance between the ``PULSE'' signal (see Figure~\ref{fig1}) in odd-numbered columns and the front-end input. When injecting a specified charge, the ``PULSE'' signal simultaneously induces approximately 310~$e^-$. This implies that the inconsistency stems from charge injection circuit rather than the front-end circuit itself. This was subsequently validated through the laser test. When evaluating only the even-numbered columns, the average threshold is measured to be 312~$e^-$ with 16.5~$e^-$ threshold dispersion, while the average ENC reaches 17.9~$e^-$.

\subsection{Laser test}

\begin{figure}
    \centering
    \subfloat[(a)]{
        \includegraphics[width=0.4\linewidth]{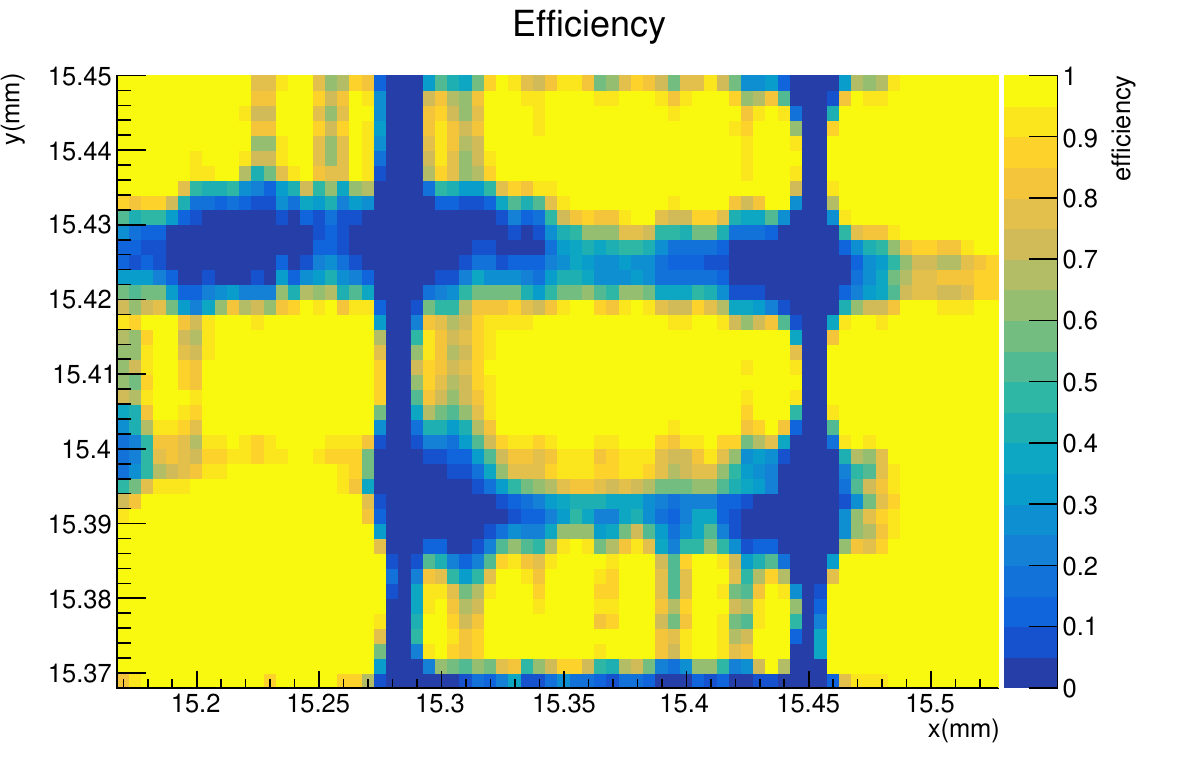} \label{Fig.8(b)}
    }
    \subfloat[(b)]{
        \includegraphics[width=0.4\linewidth]{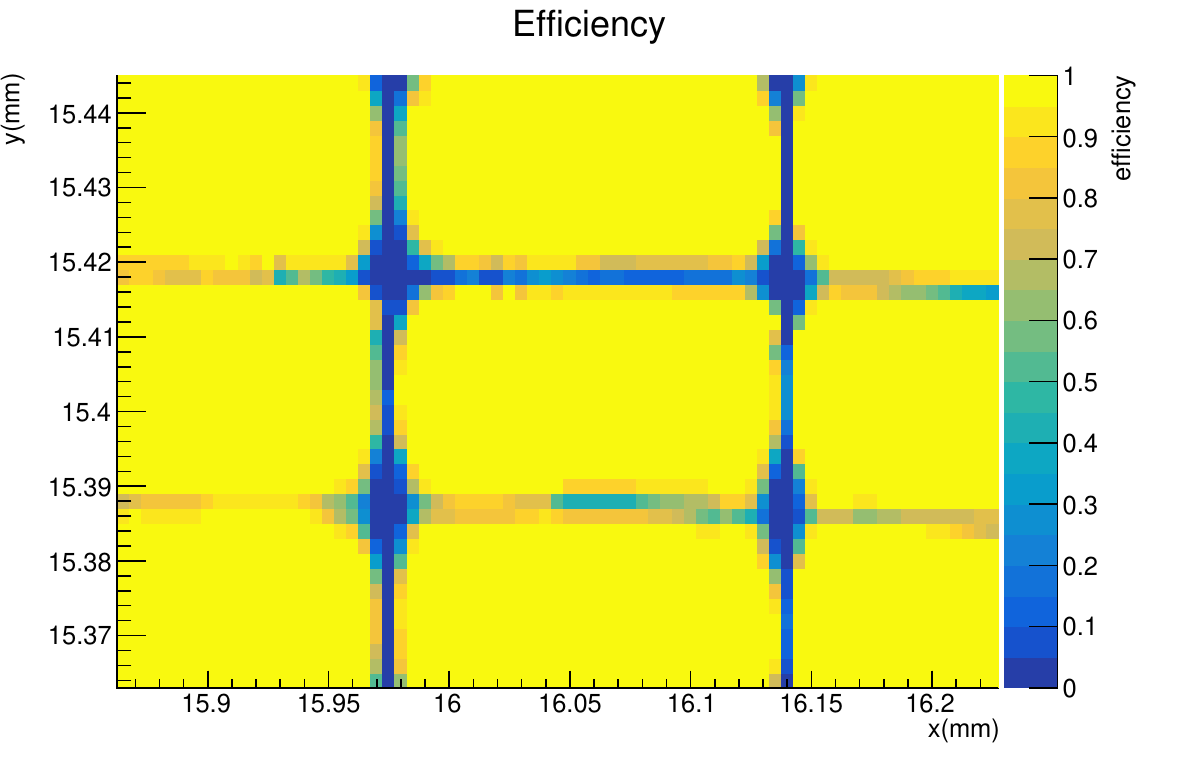}\label{Fig.8(a)}
    }
\caption{(a) Efficiency map of Sensor B, (b) Efficiency map of Sensor D}
\label{Fig.8}
\end{figure}

A preliminary laser test was conducted to verify the detection efficiency of these novel sensors. The infrared pulsed laser operates in triggering mode and a valid detection is registered when the MAPS responds within a specified time window after the trigger signal activation. The correlation between laser intensity and charge is calibrated through a S-curve method as follows. The laser spot is first positioned at the pixel center. By adjusting the laser intensity to achieve 50$\%$ detection efficiency, the corresponding intensity value is thus calibrated to the threshold charge under the same operating conditions. By scanning Sensor B and Sensor D with a laser intensity calibrated to 600~$e^-$(<0.4 MIP), Figure~\ref{Fig.8} presents their position-dependent detection efficiency distributions. The detection effciency of both sensors follows a consistent trend: near the charge-collecting electrode, both exhibit high detection efficiency of almost 100\%, which decreases toward the pixel edges and more significantly at the pixel corners (due to charge-sharing effects). However, Sensor D demonstrates notably superior performance compared to Sensor B, with efficiency drops only at minimal edge regions. This observation also aligns with simulation results. At a laser intensity equivalent to 1600~$e^-$ (about 1~MIP), both sensors achieve an average charge collection efficiency exceeding 99.9$\%$, despite their relatively high thresholds.

\section{Conclusion}
Several MAPS prototype chips featuring large-size sensors were fabricated in a 180~nm CIS process, as a detector candidate for the inner tracker of the Super Tau-Charm Facility. To satisfy the stringent material budget requirement, the MAPS design must maintain a power consumption below 100~mW/cm$^2$ while integrating timing measurements. The optimization strategy focuses primarily on reducing digital power consumption across the pixel array, achieved through a lateral sensor extension design that minimizes timestamp distribution power consumption. The front-end circuit adopts an open-loop architecture, with a threshold of 314~$e^-$ and ENC of 11.1~$e^-$ for Sensor B from simulation. Comprehensive simulations demonstrate that this architecture, when scaled to a 2~$\times$~2~cm$^2$ dimension, maintains a power consumption at 55.7~mW/cm$^2$ - well within the specified limits.

Initial prototype testing evaluated both Sensor B and Sensor D. Sensor B exhibited a mean threshold of 312~$e^-$ with 16.5~$e^-$ threshold dispersion and 17.9~$e^-$ average ENC. While Sensor D exhibited a mean threshold of 319~$e^-$ with 10.8~$e^-$ threshold dispersion and 35.0~$e^-$ average ENC. Laser-based characterization at approximately 1~MIP equivalent energy confirmed exceptional detection efficiency exceeding 99.9$\%$ for both sensor types, validating their suitability for the target application.


\section{Acknowledgments}
\label{app1}

This work was supported in part by the National Key Research and Development Program of China under Grant 2022YFA1602203, in part by the National Natural Science Foundation of China under Grant 12341502, and  the Fundamental Research Funds for the Central Universities of China (WK2360000014). We thank the Hefei Comprehensive National Science Center for their strong support on the STCF key technology research project.


 \bibliographystyle{elsarticle-num-names}\footnotesize

 \bibliography{VCI_revision_manuscript}






\end{document}